\documentclass[aps,prl,twocolumn,superscriptaddress,showpacs]{revtex4}

\usepackage{graphicx}

\newcommand{\be}{\begin{equation}}
\newcommand{\ee}{\end{equation}}

\newcommand{\Msol}{\mbox{${\rm M}_{\odot}\;$}}

\begin{document} 

\title{Thermal Evolution and Light Curves of Young Bare Strange Stars} 

\author{\surname{Dany} Page}
\affiliation{Instituto de Astronom\'{\i}a, 
             Universidad Nacional Aut\'onoma de M\'exico, 
             Mexico D.F. 04510, Mexico}
\author{\surname{Vladimir V.} Usov} 
\affiliation{Department of Condensed Matter Physics, Weizmann Institute 
             of Science, Rehovot 76100, Israel} 

\begin{abstract} 

We study numerically the cooling of a young bare strange star
and show that its thermal luminosity, mostly due to e$^+$e$^-$ 
pair production from the quark surface, may be much higher than 
the Eddington limit. 
The mean energy of photons far from the strange star is 
$\sim 10^2$~keV or even more. 
This differs both qualitatively and quantitatively from the thermal 
emission from  neutron stars and provides a definite observational 
signature for bare strange stars. 
It is shown that the energy gap of superconducting quark matter may 
be estimated from the light curves if it is in the range from 
$\sim 0.5$~MeV to a few MeV. 

\end{abstract} 



\maketitle 

Strange stars that are entirely made of deconfined quarks have
been long ago proposed as an alternative to neutron stars (e.g., 
\cite{W84}). The bulk properties (size, moment of inertia, etc.)   
of strange and neutron stars in the observed mass range $(1< M/M_\odot 
<2)$ are rather similar, and it is very difficult to discriminate 
between strange and neutron stars \cite{HZS86}. Strange quark 
matter (SQM) with a density 
of $\sim 5\times 10^{14}$ g~cm$^{-3}$ might exist 
up to the surface of a strange star \cite{AFO86}. This differs 
qualitatively from the neutron star surface and opens 
observational possibilities to distinguish strange stars from 
neutron stars. 

Normal matter (ions and electrons) with a mass 
$\Delta M\lesssim 10^{-5}M_\odot$
may be present at the SQM surface of a strange star
\cite{AFO86}. Such a massive envelope of normal matter 
could completely obscure the quark surface. 
However, a strange star at the moment of its formation is very 
hot. The temperature in the stellar interior may be as high as a 
few times $10^{11}$ K \cite{CD01,HPA91}. The rate of neutrino-induced
mass ejection from such a hot compact star is very high \cite{WB92}. 
Therefore, in a few seconds after the 
star formation the normal-matter envelope is blown away, and 
the SQM surface is nearly (or completely) 
bare \cite{U01}. A strange star remains nearly bare as long as the 
surface temperature is higher than $\sim 3\times 10^7$ K 
\cite{U97}. 

Due to the high plasma frequency of SQM, $\omega_p \sim 20$ MeV, a bare
strange star will be a very inefficient emitter of thermal X-rays as soon 
as its temperature drops below $\sim 10^{11}$ K, i.e., a few seconds 
after it is born \cite{AFO86}. This fact suggested that bare strange 
stars would be very difficult to detect.
However, the enormous surface electric field, which binds the electrons
to the quark matter, will induce intense emission
of e$^+$e$^-$ pairs \cite{U98} and subsequent hard X-ray emission, at
luminosities above the Eddington limit,
$L_{\rm Edd} \simeq 1.3 \times 10^{38}(M/M_\odot )$ erg s$^{-1}$, as
long as the surface temperature is above $\sim 5 \times 10^8$ K \cite{U01}.
This process significantly increases the possibility to detect young 
bare strange stars, but its importance depends on how long a high 
luminosity can be sustained.

We want to address this issue in this letter by modeling in detail the
thermal evolution of a young bare strange star in order to calculate 
its light curve, considering various scenarios about the state of SQM.
It should be emphasized that the resulting evolution described here will
differ both qualitatively and quantitatively from the evolution of a 
strange star with a crust, i.e., covered by a small layer of normal matter, 
\cite{SS-crust-cool},
and from the evolution of a more standard compact object \cite{PPLS00}.

Recently, it has been argued that SQM is a color superconductor
with a very high critical temperature $T_c \sim 10^{12}$ K
(for reviews, see \cite{ABR01}). 
At extremely high density the color superconductor is in the
``Color-Flavor-Locked'' (CFL) phase in which quarks of all three flavors
and three colors are paired in a single condensate. 
In this CFL phase SQM is electrically neutral and no electrons are present.
If the strange quark mass, $m_s$, is not too large the CFL phase may 
extend down to the lowest density corresponding to the surface of a
strange star, in which case the considerations of this paper are
irrelevant since no electrons would be present at the surface and
hence there would be no supercritical electric field and no e$^+$e$^-$
pair emission.
However, for sufficiently large $m_s$ the low density regime is rather
expected to be in the ``2 color-flavor SuperConductor'' (2SC) phase
in which only $u$ and $d$ quarks of two color (say $u_1$, $u_2$, $d_1$ and
$d_2$) are paired in a single condensate while the ones of the third color,
say $u_3$ and $d_3$, and the $s$ quarks of all three colors are unpaired.
In the 2SC phase electrons are present. 
We will consider strange stars made entirely of normal, i.e., unpaired, SQM, 
which may be unrealistic but is a benchmark and the case of SQM in the 
2SC phase only and with a mixture of 2SC phase with CFL phase at high density
and, finally, SQM in the 2SC phase with secondary pairing of the
$u_3$-$d_3$ and $s$ quarks with a much smaller gap.
In all cases the critical temperature is related to the energy gap 
$\Delta(0)$ at $T=0$ through $T_c \simeq 0.57 \Delta(0)$ as in BCS theory 
(in natural units $\hbar = c = k_B = 1$).

We consider, as a typical case, a 1.4 \Msol strange star which we construct 
by solving the Tolman-Oppenheimer-Volkoff equation of hydrostatic equilibrium
using an equation of state for SQM as described in \cite{HZS86,AFO86}
[with a bag constant $B = (140 {\rm MeV})^4$, QCD coupling constant
$\alpha_s \equiv g^2/4\pi = 0.3$, and $m_s = 150$ MeV].
The thermal evolution is determined by the energy conservation and
heat transport equations:
\be
C_v \frac{\partial T}{\partial t} \!=
- \frac{1}{4 \pi r^2}\frac{\partial (4 \pi r^2 F_r)}{\partial r} - Q_{\nu}
\,\,\,{\rm and}\,\,\,
F_r \!= - \kappa  \frac{\partial T}{\partial r}
\label{equ:Etran}
\ee
where $C_v$ is the specific heat of the matter,
$\kappa$ its thermal conductivity and $Q_{\nu}$ its 
neutrino emissivity, and $F_r$ is the heat flux at radius $r$.
We are actually solving the general relativistic version of these equations
with a Henyey-type cooling code \cite{P89}.
At the stellar surface, the heat flux directed outward the strange star 
is equal to the energy flux emitted from the stellar surface, 
$F_r(r=R)  =  F_\gamma + F_\pm$,
where $F_\gamma$ and $F_\pm$ are the energy flux emitted from the unit 
surface of SQM in thermal photons and $e^+e^-$ pairs, respectively. 
This and equation (\ref{equ:Etran}) give the boundary condition on 
$\partial T/\partial r$ at the stellar surface. The boundary
condition at the stellar center is $\partial T/\partial r = 0$.
In our numerical simulations, 
we adopt the values of $F_\gamma$ and $F_\pm$ from \cite{U01}.
We assume that at the initial moment, $t=0$, 
the temperature in the strange star is uniform, $T=10^{11}$~K $\sim 10$ MeV. 

The specific heat is $C_v = C_v(e) + \sum_q C_v(q)$ ($q$ running through
the nine quark components, $u_1$, ..., $s_3$) with
$C_v(i) = \frac{1}{3} \mu_i^2 T$, $\mu_i$ being the chemical potential
of the $i$th component.
When a component $i$ becomes paired its $C_v(i)$ first increases by a factor
2.426 at $T_c$ and then decreases exponentially at $T \ll T_c$
\cite{LY94a}.

The neutrino emission is due to the three quark direct Urca processes
$u_c + e^- \rightarrow d_c + \nu_e$ and 
$d_c \rightarrow u_c + e^- + \overline{\nu}_e$,
$c$ = 1, 2, 3 denoting the quark color which is not altered by weak
interactions, with emissivities \cite{I82}
\be
Q_\nu^c \simeq  3 \! \times \! 10^{25} \alpha_s  \!\!
\left[\!\frac{\mu_u \, \mu_d}{(400 {\rm MeV})^2}
          \frac{\mu_e}{10 {\rm MeV}}\!\right] \!\!
T_9^6 \, {\rm erg \, cm^{-3} \, s^{-1}}
\label{equ:DURCA}
\ee
and the corresponding, but weaker, processes with $d_c \leftrightarrow s_c$.
These processes are also strongly suppressed when one of the 
participating component is paired \cite{LY94b}.

The thermal conductivity is an essential ingredient of our
calculations since it determines how fast heat is carried to the surface
from the underlying layers.
When no quark pairing is present $\kappa$ has been calculated in \cite{HP93}:
\be
\kappa^{\rm N} \simeq
       1.7 \times 10^{21} \frac{(\mu_q/400 {\rm MeV})^2}{\alpha_s} 
       \;\, {\rm erg \; s^{-1} cm^{-1} K^{-1}}\,.
\label{equ:kappa_N} 
\ee
In the case of the 2SC phase the thermal conductivity will
be provided dominantly by the unpaired quarks $u_3$ and $d_3$ which only
suffer scattering through the exchange of the fully screened (both Debye 
and Meissner) gluon of adjoint color index 8 \cite{R00}. 
Following the method of \cite{HP93} we obtain
\be
\kappa^{\rm 2SC} \simeq
       3.5 \times 10^{22}  \frac{(\mu_q/400 {\rm MeV})^3}
       {\alpha_s^{1/2} \, T_9} 
       \;\, {\rm erg \; s^{-1} cm^{-1} K^{-1}}\,.
\label{equ:kappa_2SC}
\ee
In the CFL phase the thermal conductivity is extremely large 
\cite{SE02} because this phase is transparent to photons, 
$\kappa^{\rm CFL} \sim 10^{30} \times T_9^3 \; {\rm erg \; s^{-1} cm^{-1} K^{-1}}$.

\begin{figure}
   \includegraphics{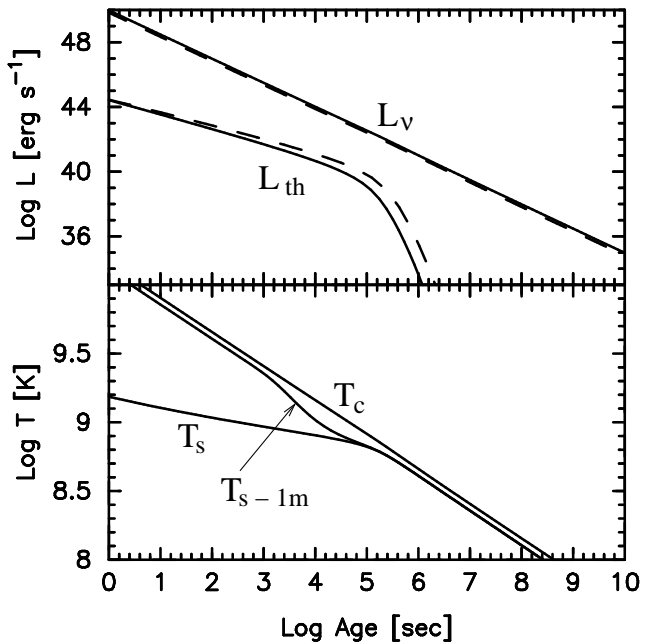}
   \caption{Upper panel: luminosities in neutrinos, $L_\nu$, and
            photons plus $e^+e^-$ pairs, $L_{\rm th}$,
            for a strange star in the normal phase (solid lines)
            and in the 2SC phase (dashed lines).
            Lower panel: temperature in the center, $T_{\rm c}$,
            at the surface, $T_{\rm s}$, and 1 meter below the
            surface, $T_{\rm s-1m}$ of the strange star in the
            normal phase. 
            The large difference between $T_{\rm s}$ and $T_{\rm s-1m}$
            at early times illustrates the enormous temperature
            gradient present just below the surface because of 
            the large $L_{\rm th}$.}
   \label{fig1} 
\end{figure}

Figure~\ref{fig1} shows the temperature and luminosities (in both neutrinos
and surface thermal emission) as a function of time $t$,
for a strange star in the normal phase and in the 2SC phase with a gap
$\Delta_{\rm 2SC} \sim 100$ MeV.
The neutrino luminosity, $L_\nu$, is always much higher than the surface
thermal luminosity, $L_{\rm th}$, i.e., neutrinos drive 
the cooling and the surface emission just follows the evolution of the 
bulk of the star. The occurrence of the 2SC phase has little 
effect on the star's evolution: the reason is that the neutrino emission 
is cut by a factor three since $u$ and $d$ quarks of 
color 1 and 2 do not contribute anymore, and the specific heat is cut by 
a factor 9/5. The neutrino cooling time scale  
is $\tau_{\rm cool} \sim C_v T/Q_\nu$ (see Eq.~(\ref{equ:Etran})), 
and it therefore does not change significantly.

We see that the thermal luminosity may be many orders of 
magnitude higher than the Eddington limit 
for a period of several days after the strange star formation.
Such a high luminosity is allowed for a bare strange 
star because at its surface SQM is bound via 
strong interaction rather than gravity and, therefore, a bare 
strange star is not subject to the Eddington limit in contrast 
to a neutron star \cite{AFO86,U98}. 
Super-Eddington luminosities are a fingerprint of hot bare strange stars. 
At $t>10^5$~s, the thermal emission decreases fast, and after about one 
month when $T_s < 3\times 10^8$ K, the thermal radiation becomes 
practically undetectable (see Fig.~\ref{fig1}).
This strong decrease of the surface emission is due to the suppression of
the e$^+$e$^-$ pairs emission by increasing degeneracy of electrons in the 
thin ''electron atmosphere''  \cite{U01}. 

Assumption that a part of the interior of the strange star is paired in the
CFL phase, with $\Delta_{\rm CFL} >$ 10 MeV,
has no effect at all on the surface thermal luminosity.
The reason is simply that it cuts down both $Q_\nu$ and $C_v$ in the same
fraction and does not affect $\tau_{\rm cool}$ at all.
We have checked it numerically with a mass of up to 1.39 \Msol of the
star, for a total mass of 1.4 \Msol, in the CFL phase and found variations
in $L_{\rm th}$ smaller than the thickness of the line in the plot
of Fig.~\ref{fig1}.
The star's temperature profile is of course affected since the CFL core
cools mostly by heat diffusion into the outer 2SC phase layer, but neutrino
emission in this outer layer is so strong that $T_s$ and $L_{\rm th}$ are
practically not affected.
This result is similar to the case of quark matter in neutron star
interiors which is undetectable if it is in the CFL phase \cite{PPLS00}.

Two effects can dramatically alter the results of Fig.~\ref{fig1}. 
They are a secondary pairing of the $u_3$ and $d_3$ quarks, and also 
possibly of the $s$ quarks, in the 2SC phase, 
and convection in the upper layers \cite{U98a}.

Pairing of the $u_3$ and $d_3$ quarks, in addition to the 2SC pairing, 
will quench the 3rd color channel of neutrino emission by the direct Urca
process of Eq.~\ref{equ:DURCA} and reduce the specific heat while pairing
of the $s$ quarks will have little effect on the neutrino emission
but will reduce $C_v$.
The suppression of $Q_\nu$ and $C_v$ is very strong only if the resulting 
gap is nodeless, while a gap with nodes on the Fermi surface will only 
produces a reduction of the order of 
$(T/T_c)^2$ if nodes are at isolated points and of order of 
$T/T_c$ if nodes are 1D lines.
The cooling time scale, $\tau_{\rm cool}$, can be increased or reduced
depending on whether $Q_\nu$ or $C_v$ is the most strongly suppressed.
Since very little is known on these possible secondary gaps 
(see however \cite{ARW98}) we consider three cases for the $u_3$-$d_3$ gap 
$\Delta_3$ and the $s$ gap $\Delta_s$:
[A] $\Delta_3 = 0.3$ Mev and $\Delta_s = 3$ MeV,
[B] $\Delta_3 = \Delta_s = 3$ MeV, and
[C] $\Delta_3 = 3.0$ Mev and $\Delta_s = 0.3$ MeV.
These three cases span a large range of suppression of $Q_\nu$ and/or $C_v$.

\begin{figure}[t]
   \includegraphics{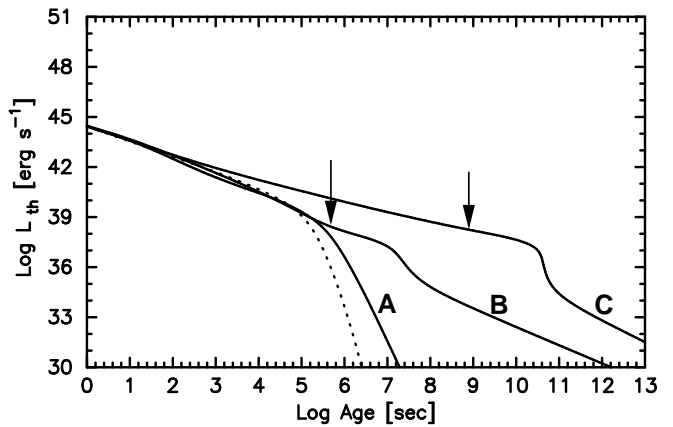}
   \caption{The thermal luminosity of the strange star
            in the 2SC phase with secondary
            pairing of the $u_3$-$d_3$ and $s$ quarks for three different
            scenarios of pairing [A], [B] and [C] as described in the text
            (solid lines).
            Arrows on curves [B] and [C] point the time at which
            $L_\nu$ becomes lower than $L_{\rm th}$; 
            in case [A] $L_\nu \gg L_{\rm th}$ at all times.
	    The dotted line shows the evolution of $L_{\rm th}$ without
	    any pairing for comparison (from Fig.~\protect\ref{fig1}).}
   \label{fig2} 
\end{figure}

Figure~\ref{fig2} shows the resulting range of $L_{\rm th}$ possibly attained 
through this secondary pairing. We assumed nodeless gaps to maximize the 
effects, but have verified explicitly that gaps with nodes give results 
which are intermediates between the three cases presented here.
Case [A] has very little $Q_\nu$ suppression and only partial reduction
of $C_v$, and it naturally differs little from the case of no pairing.
Cases [B] and [C] have strong $Q_\nu$ suppression and their cooling is
eventually driven by the surface thermal emission.
Since case [B] also has a strong $C_v$ reduction it cools faster than case [C] 
whose $C_v$ is more moderately suppressed.
The knees on the curves [B] and [C] at times $t \sim 10^7$ sec
and $t \sim 3\times 10^{10}$ sec, respectively, 
correspond to the moment when 
the star becomes almost isothermal.

In all the models presented above, at early times, the temperature gradient 
just below the surface (see Fig.~\ref{fig1}, lower panel) is  
many orders of magnitude higher than the adiabatic temperature gradient 
\cite{GY94}
\be
\left. \frac{dT}{dr}\right|_{\rm ad} \cong \frac{T}{3 n_q} \frac{d n_q}{d r}
= \frac{T}{\mu_q} \frac{d \mu_q}{d r}\sim 300 \times T_9\,\,\,
{\rm K}\,{\rm cm}^{-1}
\label{equ:dTdr_ad}
\ee
where $n_q$ is the quark number density.
If convection can develop, given the shallowness of the superadiabatic
layer we can, as a first approximation, consider the
star as having a uniform temperature.
Moreover, in the case of pairing of the $u_3$-$d_3$ quarks, and/or $s$ quarks,
we may expect convective counterflow of the $u_3$-$d_3$, and/or $s$,
superfluid which may be even more efficient that convection to erase any 
temperature gradient. 

\begin{figure} 
   \includegraphics{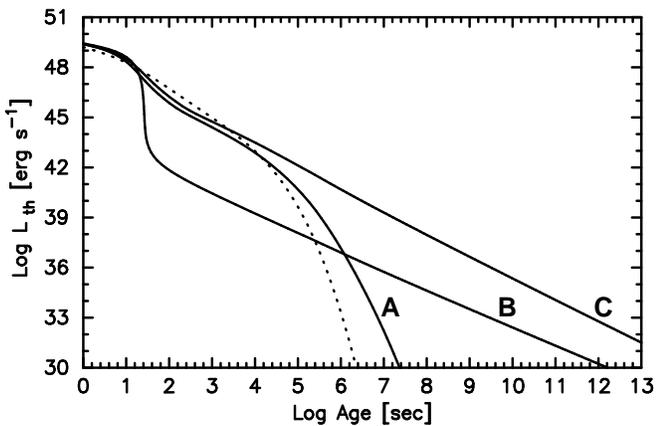}
   \caption{The thermal luminosity for the same scenarios as in 
            Fig.~\protect\ref{fig2}. 
            It is assumed that the star is isothermal due to either 
            convection or superfluid counterflow.
            The dotted curve shows the evolution of $L_{\rm th}$
            without any pairing for comparison.}
   \label{fig3} 
\end{figure}

Figure~\ref{fig3} shows the resulting range of $L_{\rm th}$ for isothermal 
stars for the same three cases of secondary pairing as in Fig.~\ref{fig2}.
In cases [B] and [C], a few seconds after the star
formation the $u_3$-$d_3$ pairing occurs
and subsequently $L_\nu < L_{\rm th}$ 
while in case [A] neutrino losses drive the cooling during all times.
The sharp drop of $L_{\rm th}$ in case [B] at $t \sim 20$ s is due to the 
strong suppression of $C_v$ from pairing and does not occur in cases [A] and
 [C] in which the $u_3$-$d_3$ or $s$ gaps are smaller.
At early times the isothermal models have a much higher thermal luminosity
that the diffusive ones of Fig.~\ref{fig2} since their surface temperature is
much higher.
However, the isothermal models whose cooling becomes driven by thermal 
emission, [B] and [C], cool faster and eventually have a lower 
$L_{\rm th}$ than the diffusive ones.
Naturally, isothermal and diffusive models follows the same evolution once
the latter become isothermal, i.e., after the knees of cases [B] and [C]
in Fig.~\ref{fig2}.

How much the density dependence of the chemical composition of SQM,
and of the pairing gaps, may reduce convection, and superfluid counterflow, 
is an open question and precludes us to firmly decide which
cooling trajectories, Fig.~\ref{fig2} or \ref{fig3}, are the correct ones.
However, the gradients of both $\mu_i$'s and $\Delta$'s are small in SQM
and one may expect only small reductions, i.e., the isothermal models of 
Fig.~\ref{fig3} are probably more appropriate that the diffusive ones of 
Fig.~\ref{fig2}.

In our simulations we assumed that neutrinos escape freely from 
the stellar interior. This is valid only in a few seconds after the 
strange star formation when the internal 
temperature is less than $\sim 10^{10}$~K \cite{HPA91}. 
In this case, $e^+e^-$ pairs created at the surface of SQM 
prevail over photons in the surface thermal emission \cite{U01}
but pairs outflowing from the stellar surface mostly annihilate 
into photons in the vicinity of the strange star \cite{U01,AMU}. 

In the process of the star cooling the photon spectrum varies 
significantly. At very high luminosities, 
$L_{\rm th}> 10^{43}$~ergs~s$^{-1}$, we expect that 
the photon spectrum is nearly blackbody with a temperature 
$T_{\rm BB} \simeq T_0(L_{\rm th}/ 10^{43}~{\rm erg~s}^{-1})^{1/4}$, where 
$T_0\simeq 2\times 10^8$~K \cite{P90}. For intermediate luminosities, 
$10^{41}<L_{\rm th}<10^{43}$~erg~s$^{-1}$, the effective 
temperature of photons is more or less constant, 
$T_{\rm BB} \sim T_0$ \cite{AMU}. 
At $L_{\rm th}< 10^{41}$ erg~s$^{-1}$, the photon spectrum essentially
differs from the blackbody spectrum, and its 
hardness increases when $L_{\rm th}$ decreases. 
This is because photons that form in annihilation of 
$e^+e^-$ pairs do not have enough time for thermalization 
before they escape from the strange star vicinity. 
When the photon luminosity decreases from 
$\sim 10^{41}$ erg~s$^{-1}$ to $\sim 10^{36}$ erg~s$^{-1}$, 
the mean energy of photons increases from $\sim 30$~keV 
to $\sim 500$~keV while the spectrum 
of photons eventually changes into a very wide $(\Delta E 
/E\sim 0.3)$ annihilation line of energy $E\sim 500$~keV \cite{AMU}. 
Such a variability of the photon spectrum together with the 
light curves calculated in this paper could be 
a good observational signature of a young bare strange star. 


\begin{acknowledgments}
VVU wants to thank the Institute of Astronomy, UNAM, Mexico, where 
this work was carried out, for its wonderful hospitality. 
This work was supported by a grant from the Mexican Conacyt (\# 27987-E).
\end{acknowledgments}

 

\end{document}